\documentclass[]{article}

\usepackage{savesym}
\usepackage{graphicx}

\usepackage{authblk}
\usepackage{lineno}

\savesymbol{tablenum}
\usepackage{siunitx}
\newcommand\unit[3][]{\SI[#1]{#2}{#3}}
\restoresymbol{SIX}{tablenum}

\usepackage[version=3]{mhchem}
\usepackage{makecell}

\usepackage{geometry}
\usepackage[utf8]{inputenc}       
\usepackage[T1]{fontenc}          

\usepackage{amsmath}
\usepackage{amssymb}
\usepackage{mathrsfs}             

\usepackage{lineno}

\usepackage{tikz}
\usetikzlibrary{plotmarks,patterns,calc,decorations.markings}

\usepackage{booktabs}

\catcode`\@=11 
\def\parenbar{\mathpalette\p@renb@r}
\def\p@renb@r#1#2{\vbox{%
\ifx#1\scriptscriptstyle \dimen@.7em\dimen@ii.2em\else
\ifx#1\scriptstyle \dimen@.8em\dimen@ii.25em\else
\dimen@1em\dimen@ii.4em\fi\fi \offinterlineskip
\ialign{\hfill##\hfill\cr
\vbox{\hrule width\dimen@ii}\cr
\noalign{\vskip-.3ex}%
\hbox to\dimen@{$\mathchar300\hfil\mathchar301$}\cr
\noalign{\vskip-.3ex}%
$#1#2$\cr}}}
\catcode`\@=12 
\def\nuan{\parenbar{\nu}\kern-0.4ex}

\newlength{\smfigwidth}
\newlength{\figwidth}
\newlength{\captwidth}
\usepackage[pdflatex]{trimclip}

\usepackage{hyperref}
\usepackage[all]{hypcap}          

\title{ANTARES neutrino search for time and space correlations with IceCube high-energy neutrino events}

\date{}

\author[1]{A.~Albert}
\author[2]{M.~Andr\'e}
\author[3]{M.~Anghinolfi}
\author[4]{G.~Anton}
\author[5]{M.~Ardid}
\author[6]{J.-J.~Aubert}
\author[7]{J.~Aublin}
\author[7]{B.~Baret}
\author[8]{J.~Barrios-Mart\'{\i}}
\author[9]{S.~Basa}
% ...

\affil[1]{\scriptsize{Universit\'e de Strasbourg, CNRS,  IPHC UMR 7178, F-67000 Strasbourg, France}}
\affil[2]{\scriptsize{Technical University of Catalonia, Laboratory of Applied Bioacoustics, Rambla Exposici\'o, 08800 Vilanova i la Geltr\'u, Barcelona, Spain}}
\affil[3]{\scriptsize{INFN - Sezione di Genova, Via Dodecaneso 33, 16146 Genova, Italy}}
\affil[4]{\scriptsize{Friedrich-Alexander-Universit\"at Erlangen-N\"urnberg, Erlangen Centre for Astroparticle Physics, Erwin-Rommel-Str. 1, 91058 Erlangen, Germany}}
\affil[5]{\scriptsize{Institut d'Investigaci\'o per a la Gesti\'o Integrada de les Zones Costaneres (IGIC) - Universitat Polit\`ecnica de Val\`encia. C/  Paranimf 1, 46730 Gandia, Spain}}
\affil[6]{\scriptsize{Aix Marseille Univ, CNRS/IN2P3, CPPM, Marseille, France}}
\affil[7]{\scriptsize{APC, Univ Paris Diderot, CNRS/IN2P3, CEA/Irfu, Obs de Paris, Sorbonne Paris Cit\'e, France}}
\affil[8]{\scriptsize{IFIC - Instituto de F\'isica Corpuscular (CSIC - Universitat de Val\`encia) c/ Catedr\'atico Jos\'e Beltr\'an, 2 E-46980 Paterna, Valencia, Spain}}
\affil[9]{\scriptsize{LAM - Laboratoire d'Astrophysique de Marseille, P\^ole de l'\'Etoile Site de Ch\^ateau-Gombert, rue Fr\'ed\'eric Joliot-Curie 38,  13388 Marseille Cedex 13, France}}
% ...

\begin{document}

\maketitle

\begin{abstract}

In the past years, the IceCube Collaboration has reported in several analyses the observation of astrophysical high-energy neutrino events. Despite a compelling evidence for the first identification of a neutrino source, TXS 0506+056, the origin of the majority of these events is still unknown.
In this paper, a possible transient origin of the IceCube astrophysical events is searched for using neutrino events detected by the ANTARES telescope. The arrival time and direction of 6894 track-like and 160 shower-like events detected over 2346 days of livetime are examined to search for coincidences with 54 IceCube high-energy track-like neutrino events, by means of a maximum likelihood method. No significant correlation is observed and upper limits on the one-flavour neutrino fluence from the direction of the IceCube candidates are derived. 
The non-observation of time and space correlation within the time window of 0.1 days with the two most energetic IceCube events constrains the spectral index of a possible point-like transient neutrino source, to be harder than $-2.3$ and $-2.4$ for each event, respectively.

\end{abstract}

\maketitle

\section{Introduction}\label{sec:INTRO}

The observation of high-energy astrophysical neutrinos reported by the IceCube Collaboration in the last few years represents a crucial step forward in the field of neutrino astronomy and strongly motivates independent searches for their origin.
The first significant evidence of a diffuse flux of extraterrestrial neutrinos in the TeV-PeV range was observed in the ``High-Energy Starting Events" (HESE) sample of IceCube \cite{IC3years, IC4yproc, IC6yproc}. 
The spectral energy distribution of the 82 events recorded in 6 years of data taking is described in \cite{IC6yproc} as a single power law: $E_{\nu}^2\Phi(E) = 2.46 \pm 0.8 \times 10^{-8} (E_{\nu}/100$ TeV$)^{-0.92^{+0.33}_{-0.29}}$ GeV cm$^{-2}$ s$^{-1}$ sr$^{-1}$.
The ANTARES Collaboration has investigated the possibility that this signal partially originates from point-like steady sources in a wide region close to the Galactic Centre and from the position of 13 HESE reconstructed as tracks \cite{PS}. Since no significant excess was observed, strong constraints on Galactic steady-source contributions of the HESE sample were set. 
\newline Another recent measurement by IceCube of the cosmic neutrino flux is based on the analysis of eight years of track-like events from the Northern Hemisphere \cite{IC-VHE-29, IC-VHE-36}, hereafter referred to as ``the Muon sample''. The analysis of the 36 muon neutrino events with reconstructed energy $> 200$ TeV selected by IceCube resulted in a best-fit of the astrophysical spectrum given by a single power law: $E_{\nu}^2\Phi(E) = 1.01^{+0.26}_{-0.23} \times 10^{-8} (E_{\nu}/100$ TeV$)^{-0.19 \pm 0.10}$ GeV cm$^{-2}$ s$^{-1}$ sr$^{-1}$ \cite{IC-VHE-36}. This result presents some tensions with the HESE measurement, which comes from the all-sky analysis, dominated by shower-like events. The proposed hypothesis of the diffuse Galactic neutrino emission being a possible cause of the discrepancy \cite{ICanomaly} has been severely constrained by both ANTARES and IceCube \cite{GCDiffuse, GCDiffuse2}. \newline
Recently, a high-energy neutrino detected by IceCube was found to be positionally coincident with the direction of a known blazar, TXS 0506+056, in a state of enhanced activity observed in $\gamma$-rays and at other wavelengths of the electromagnetic spectrum \cite{ICmultimessenger}. 
Moreover, an a posteriori time-variability study of the neutrino emission revealed a flare that occured in 2014/2015 \cite{ICTXS}. An ANTARES follow-up triggered by this result yielded no significant observation for neutrinos in space and/or time correlation either with the high-energy event or with the 2014/2015 flare \cite{ANTARESTXS}. 
\newline Finally, the observation of two spatially compatible events from the HESE sample with a time difference of less than one day, with a p-value of 1.6\% \cite{LightHouseSgrA}, could be interpreted as the signature of another flaring source. All these results reinforce the motivation of a time correlation study between ANTARES and IceCube events. Such a correlation would support the hypothesis of the IceCube events being originated from flaring episodes.

In the analysis presented in this paper, a total of 54 candidate cosmic neutrino events are selected from the IceCube HESE and Muon samples and treated as potential transient neutrino sources. Only events classified as muon tracks, lying within the ANTARES field-of-view and provided with an angular error, are included in the list. In case of events that are present in both samples, only the one with smaller angular uncertainty is considered. The equatorial coordinates, observation time and median angular error of the selected HESE candidates numbered 1 to 37 are extracted from \cite{IC3years}, 38 to 54 from \cite{IC4yproc} and 55 to 82 from \cite{IC6yproc}; Muon events numbered 1 to 29 are extracted from \cite{IC-VHE-29} and 30 to 36 from \cite{IC-VHE-36}. All information is reported in Tables~\ref{tab:candidatesHESE} and~\ref{tab:candidatesMuon}. The angular uncertainty corresponds to the median angular error reported by the IceCube Collaboration in the case of the HESE sample. For events from the Muon sample, an estimation of the median angular uncertainty is derived from the angular errors on the equatorial coordinates provided by the IceCube Collaboration. This is done assuming that the median angular errors on the declination, $\delta$, and the right ascension, $\alpha$, follow independent Gaussian distributions with standard deviation given by the angular errors. The standard deviation of the two-dimensional Gaussian, function of $\delta$ and $\alpha$, is then employed as median angular uncertainty. In this sample, a conservative minimum value of $\unit{1}{\degree}$ for the angular uncertainty is assumed. 
\newline In contrast to time-integrated searches, the information of the neutrino arrival times is exploited to enhance the discovery potential. When dealing with transient emissions, the background of atmospheric neutrinos can be significantly reduced by limiting the search to a small time window around the source flare. In this work, a maximum likelihood approach is followed to look for spatial and temporal coincidences between the selected ANTARES events and the IceCube HESE and Muon candidates. 

The paper is organised as follows. In Section \ref{sec:ANTARES}, the ANTARES neutrino telescope and the data sample are described. The search method is explained in Section \ref{sec:METHOD}. In Section \ref{sec:RESULTS}, the results of the analysis are presented, while the conclusions are summarized in Section \ref{sec:CONCL}.

\section{ANTARES neutrino telescope and data sample}\label{sec:ANTARES}

The ANTARES neutrino telescope \cite{antares} is located 40 km off-shore from Toulon, France, anchored 2475 m below the surface of the Mediterranean Sea. A three-dimensional array of 885 photomultiplier tubes (PMTs) detects the Cherenkov light induced by charged particles produced in neutrino interactions within and around the instrumented volume. The 10-inch PMTs, distributed along 12, 450 m long, vertical lines, face \unit{45}{\degree} downward in order to optimise the detection of light from upgoing particles. The position, time and collected charge of the signals in the PMTs ($\it{hits}$) are used to infer the direction and energy of the incident neutrino. \newline
Two event topologies can be identified in the ANTARES neutrino telescope: track-like and shower-like. The former can be the signature of a long-range muon produced in charged current (CC) interactions of muon neutrinos in the proximity of the detector. For this event topology, the direction of the parent neutrino can be reconstructed with a median angular resolution of \unit{0.4}{\degree} \cite{lastPS}. Shower-like events are mainly induced by neutral current (NC) interactions, and by ${\nu_e}$ and ${\nu_\tau}$ CC interactions. Since the shower elongation is of a few metres, the whole shower appears as a point-like light source in the ANTARES detector. A median angular resolution of about \unit{3}{\degree} can be achieved for high-quality selected events \cite{TANTRA}. The analysis presented in this paper includes both track-like and shower-like events recorded in ANTARES between the 1st of December 2008 and the 31st of December 2016 for a total livetime of 2346 days, covering the whole considered IceCube observation time (6 years and 8 years for the HESE and Muon samples, respectively).
The events are selected following the chain of cuts defined in the latest ANTARES point-like source analysis \cite{PS}. A summary of the different selection criteria for tracks and showers is given below.

\emph{Track Selection.} The selection of muon-neutrino-induced events is optimised using parameters provided by the track reconstruction algorithm -- a multi-step fitting procedure that estimates the direction and the position of the muon by means of a maximum likelihood method \cite{antPS4y}. Cuts are applied on the reconstructed zenith angle ($\cos\theta_{tr} > -0.1$), the estimated angular error ($\beta_{tr} < \unit{1}{\degree}$) and the parameter that describes the quality of the reconstruction ($\Lambda > -5.2$) in order to increase both the quality and the purity of the neutrino sample. 
Further cuts, described in \cite{PS}, are imposed on energy-related variables in order to guarantee the validity of the muon energy estimator employed in the analysis. For each event, a proxy for the muon energy is provided based on the parameter $\rho$ \cite{dEdXcuts,dEdX} which computes the muon energy deposition per unit path length. 
A total of 6894 neutrino candidates are selected in the track channel, with an expected atmospheric muon contamination of 13\%.

\emph{Shower Selection.} Only events not selected as tracks are considered in the shower channel.  
Showers are selected if reconstructed as upgoing or coming from close to the horizon ($\cos\theta_{sh} > -0.1$) with constraints on the angular error estimate ($\beta_{sh} < \unit{30}{\degree}$) and on the interaction vertex, which is required to lie within a fiducial volume slightly larger than the instrumented volume. Additional selection cuts based on parameters provided by two different shower reconstruction algorithms are applied to further reduce the remaining background from mis-reconstructed atmospheric muons. A detailed description of these cuts can be found in \cite{PS}. The selection yields a total of 160 neutrino candidates in the shower channel, with an estimated 43\% of atmospheric muon contamination.

\section{Analysis method and expected performances}\label{sec:METHOD}

The directions of the 54 IceCube candidates are investigated to search for spatial and temporal clustering of events above the known background expectation following a maximum likelihood ratio approach. The likelihood describes the ANTARES data in terms of signal and background probability density functions (PDFs) and is defined as:

\begin{align} \label{eq:lik}
    \log \mathcal L_\mathrm{s+b} = \sum_{\mathcal J\in{\{\it{tr}, \it{sh}\}}} \sum_{i\in\mathcal J} \log \Big[ \mu_\mathrm{sig}^{\mathcal J}
       {\mathcal S}^{\mathcal J}_{i} + {\mathcal N}^{\mathcal J}  {\mathcal B}^{\mathcal J}_{i}\Big]
    - \mu_\mathrm{sig} ,
\end{align} 

\noindent where ${\mathcal S}^{\mathcal J}_{i}$ and ${\mathcal B}^{\mathcal J}_{i}$ are the values of the signal and background PDFs for the event $i$ in the sample ${\mathcal J}$ ($\it{tr}$ for tracks, $\it{sh}$ for showers), while $\mu_\mathrm{sig}^{\mathcal J}$ and ${\mathcal N}^{\mathcal J}$ are respectively the number of unknown signal events and the total number of data events in the $\mathcal J$ sample.
The combined information of three parameters -- direction, observation time and energy -- is included in the definition of the PDFs in order to enhance the signal-to-background discrimination. While atmospheric neutrino events are rather randomly distributed, neutrinos from transient point-like sources are expected to accumulate around the source position and flaring time, i.e.\ the direction $(\alpha, \delta)$ and the observation time $t_{IC}$ of the considered IceCube candidate. The energy information helps to distinguish signal from background, as a softer energy spectrum is predicted for atmospheric neutrinos ($E^{-\gamma}$ with $\gamma \sim -3.6$ \cite{dEdX}) with respect to the expected signal.
Slightly different definitions of the PDFs are used in the track and in the shower channels. For each track-like event $i$, the probability of being reconstructed as signal or background is given by:

\begin{align}
\begin{split}\label{eq:sgPDF}
\mathcal {S}^{\it tr}_{i}  =  \mathcal {S}^{\rm space}(\Delta\Psi_i, \beta_i | \gamma) \cdot \mathcal {S}^{\rm energy}(\rho_i, \beta_i | \delta_i, \gamma) \cdot \mathcal {S}^{\rm time}(t_i) , 
\end{split}\\
\begin{split}\label{eq:bgPDF}
\mathcal {B}^{\it tr}_{i}  =  \mathcal {B}^{\rm space}(\delta_i) \cdot \mathcal {B}^{\rm energy}(\rho_i, \beta_i | \delta_i) \cdot \mathcal {B}^{\rm time}(t_i) .
\end{split}
\end{align}
As for the shower-like events, the probabilities are computed as

\begin{align}
\begin{split}\label{eq:sgPDF}
\mathcal {S}^{\it sh}_{i}  =  \mathcal {S}^{\rm space}(\Delta\Psi_i | \gamma) \cdot \mathcal {S}^{\rm energy}(N^{\rm hits}_{i} | \gamma) \cdot \mathcal {S}^{\rm time}(t_i) , 
\end{split}\\
\begin{split}\label{eq:bgPDF}
\mathcal {B}^{\it sh}_{i}  =  \mathcal {B}^{\rm space}(\delta_i) \cdot \mathcal {B}^{\rm energy}(N^{\rm hits}_{i}) \cdot \mathcal {B}^{\rm time}(t_i),
\end{split}
\end{align}

where:

\begin{itemize}

     \item $\mathcal {S}^{\rm space}$ is a parameterization of the Point Spread Function (PSF), i.e. the probability density function of reconstructing an ANTARES event $i$ at a given angular distance $\Delta\Psi_i$ from the true source location, i.e.\ the position of the IceCube candidate. The shape of the PSF is determined from Monte Carlo simulations of cosmic neutrinos assuming a $E^{-\gamma}$ energy dependence of the spectrum with variable spectral index $\gamma$, which is fitted in the likelihood maximisation. In the case of the track channel, the information of the event angular error estimate $\beta_{i}$ is also included.

	\item $\mathcal {B}^{\rm space}$ yields the probability of reconstructing a background event at a certain declination  $\delta_i$. It is derived from data using the observed declination distribution of the selected events.

    \item $\mathcal {S}^{\rm energy}$ and $\mathcal {B}^{\rm energy}$ give the probability for a signal or background event to be reconstructed with a certain value of the energy related parameter ($\rho$ for tracks and the number of hits used by the reconstruction algorithm, $N^{\rm hits}$, for showers). Monte Carlo simulations of $E^{-\gamma}$ energy spectrum cosmic neutrinos (signal) assuming neutrino flavour equipartition at Earth and of atmospheric neutrinos using the spectrum of \cite{atmonu} (background) are used to derive the energy PDFs. In the track channel, the information of the event angular error estimate $\beta_{i}$ is also considered and the dependence of the energy estimator on the declination $\delta_i$ of the event is taken into account by generating both PDFs in steps of 0.2 in $\sin\delta$.

    \item $\mathcal {S}^{\rm time}$ is the signal time-dependent PDF. In this analysis, a generic Gaussian time profile for the signal emission is assumed, $\mathcal{S}^{\rm time}(t_i) = \frac{1}{\sqrt{2\pi}\sigma_t}e^{(-\frac{(t_i - t_{IC})^2}{2\sigma_t^2})}$,   with $t_i$ being the detection time of the ANTARES event $i$, $t_{IC}$ the observation time of the considered IceCube candidate, and $\sigma_t$ the unknown flare duration, fitted in the likelihood maximisation.
    
    \item $\mathcal {B}^{\rm time}$ describes the probability to observe a background event at a given time $t_i$.  Given the small expected contribution of a cosmic signal in the overall data set, this PDF is built using the time distribution of data events, ensuring a time profile proportional to the measured data. To avoid statistical fluctuations, this PDF is computed applying looser selection criteria than those of the final sample.

\end{itemize}

The likelihood of equation~(\ref{eq:lik}) is maximised independently at the position of each IceCube event leaving as free parameters the number of signal events $\mu_\mathrm{sig} = {\mu}^{tr}_\mathrm{sig} + {\mu}^{sh}_\mathrm{sig}$, the signal spectral index $\gamma$ and the flare duration $\sigma_t$, providing the best-fit values $\hat{\mu}_\mathrm{sig}$, $\hat{\gamma}$, $\hat{\sigma}_t$ for each investigated source. Moreover, the position in the sky of the fitted source is left free to vary around the position of the IceCube event within a cone with opening angle twice as large as its angular uncertainty. In the maximisation, the value of the spectral index can range between 1.5 and 3.5, while values between 0.1 and 120 days are allowed for the flare duration. The lowest precision of the observation time of the IceCube candidates provided by the IceCube Collaboration sets the lower bound to 0.1 days, while the choice of 120 days as upper bound is imposed by the fact that the time distance between the last recorded IceCube candidate (HESE ID 82) and the last ANTARES available fully-calibrated data is $\sim 240$ days. Thus, more than 95\% of the signal events from a Gaussian flare with $\sigma = 120$ days and centered at the observation time of HESE ID 82 could be detected within the considered ANTARES data taking period.

\begin{figure*}[ht]
	\centering
    \includegraphics[width=0.48\linewidth]{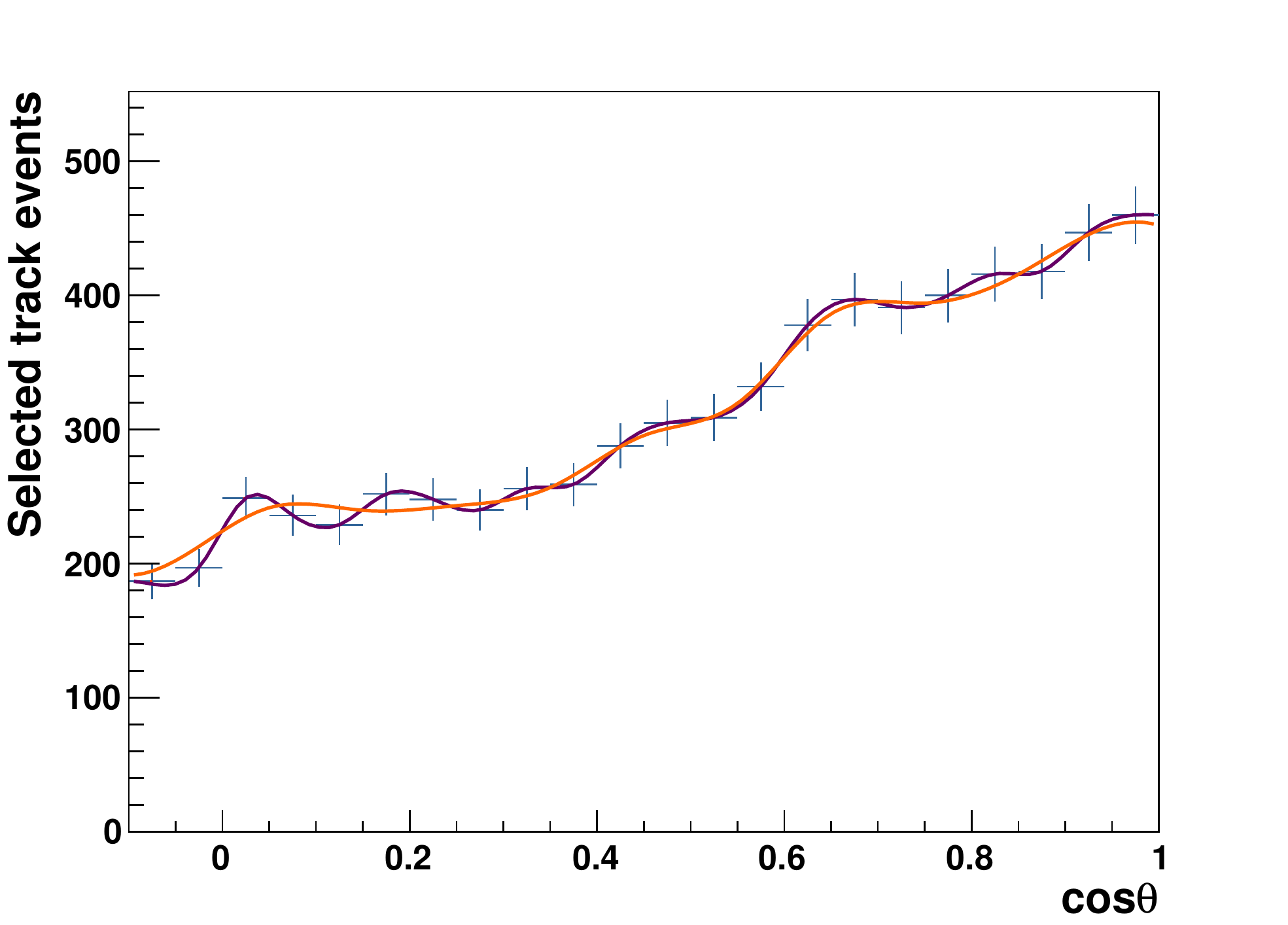}
    \includegraphics[width=0.48\linewidth]{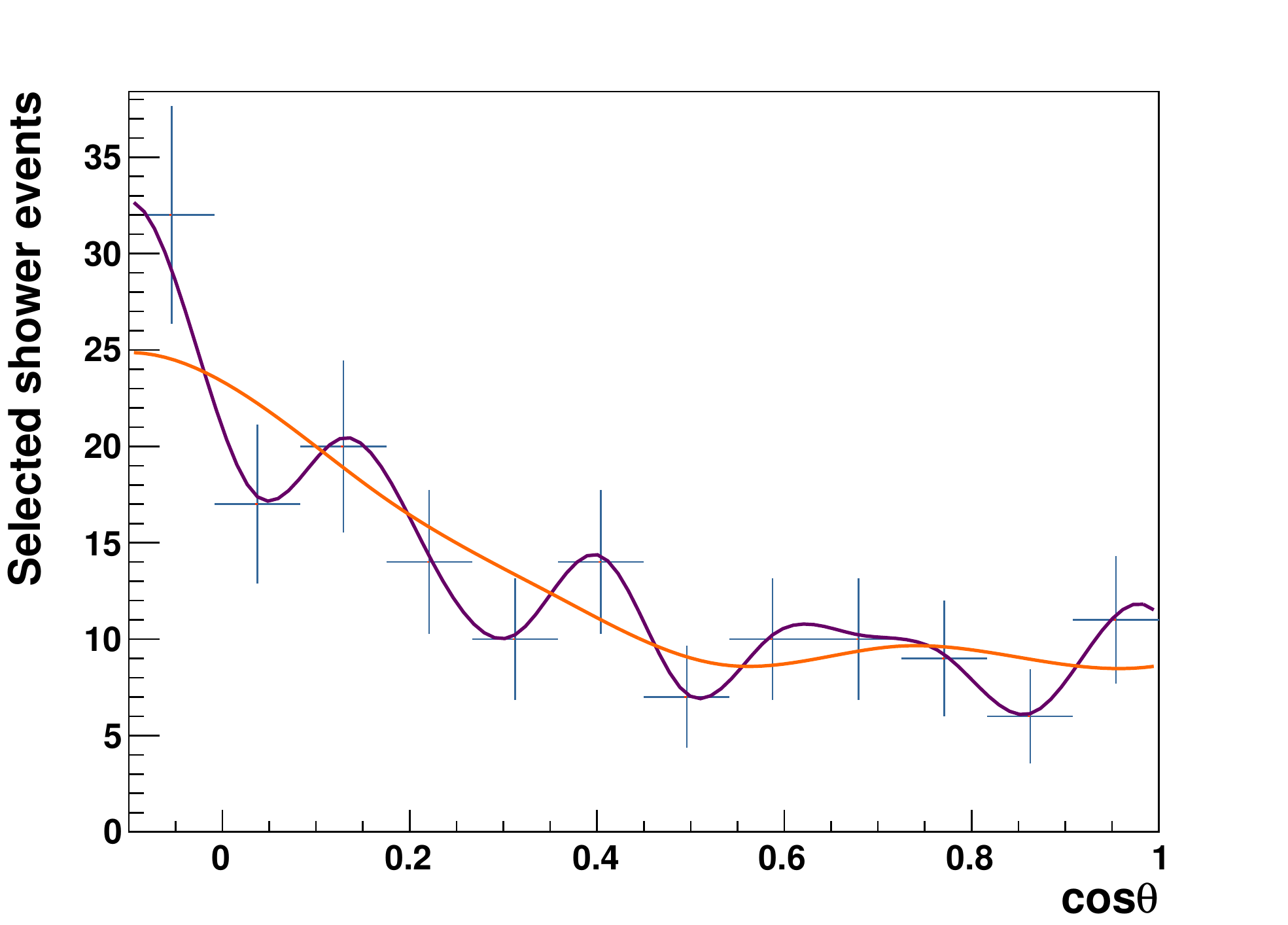}      
    \caption{Number of selected track-like (left) and shower-like (right) data events collected in 2346 days of livetime as a function of the reconstructed zenith angle. The spline functions, $P(\theta)$ and $O(\theta)$, are shown as purple and orange lines.}
    \label{fig:bgrate}
\end{figure*}

The significance of any cluster of ANTARES events around an IceCube candidate is determined by a test statistic $\mathcal Q$ derived from the likelihood as

\begin{equation}
    \mathcal Q = \log \mathcal L_\mathrm{s+b} - \log \mathcal L_\mathrm{b} ,
    \label{eq:teststat}
\end{equation}
where $\log \mathcal L_\mathrm{s+b}$ is the likelihood defined in equation~(\ref{eq:lik}) evaluated with the best-fit values ($\mu_\mathrm{sig} = \hat{\mu}_\mathrm{sig}$, $\sigma_t = \hat{\sigma}_t$, $\gamma = \hat{\gamma}$, $\alpha = \hat{\alpha}$, $\delta = \hat{\delta}$) and $\log \mathcal L_\mathrm{b}$ is the likelihood  evaluated in the background-only case ($\mu_\mathrm{sig} = 0$). 
The $\mathcal Q$ distributions for different signal strengths are determined from pseudo-experiments (PEs), i.e.\ performing the search for time and spatial correlation on scrambled data. In each PE, a fake sky-map containing a known number of signal events injected into a background-only dataset is generated. The simulated directions and times of the background events are randomly drawn from the zenith, azimuth and time distributions as seen in the actual data. The distribution of the reconstructed zenith angle is parametrised by two different spline functions, $P(\theta)$ and $O(\theta)$, shown in Figure \ref{fig:bgrate}. In order to account for possible systematic uncertainties on the background, the zenith-dependent distribution of background events, $\mathcal Z(\theta)$, in each PE is taken as $\mathcal Z(\theta) = P(\theta) + r \cdot ( O(\theta) - P(\theta) )$, with $r$ being a random number drawn from a uniform distribution between -1 and 1.
The simulated signal events are injected around a given candidate position assuming an unbroken power-law $E^{-\gamma}$ energy spectrum with $\gamma$ being the tested spectral index.
A random time drawn from a Gaussian distribution characterized by a mean and a standard deviation given by the IceCube candidate observation time and the tested flare duration is assigned to each signal event.

\begin{figure}[h]
    \centering
    \includegraphics[width=0.65\linewidth]{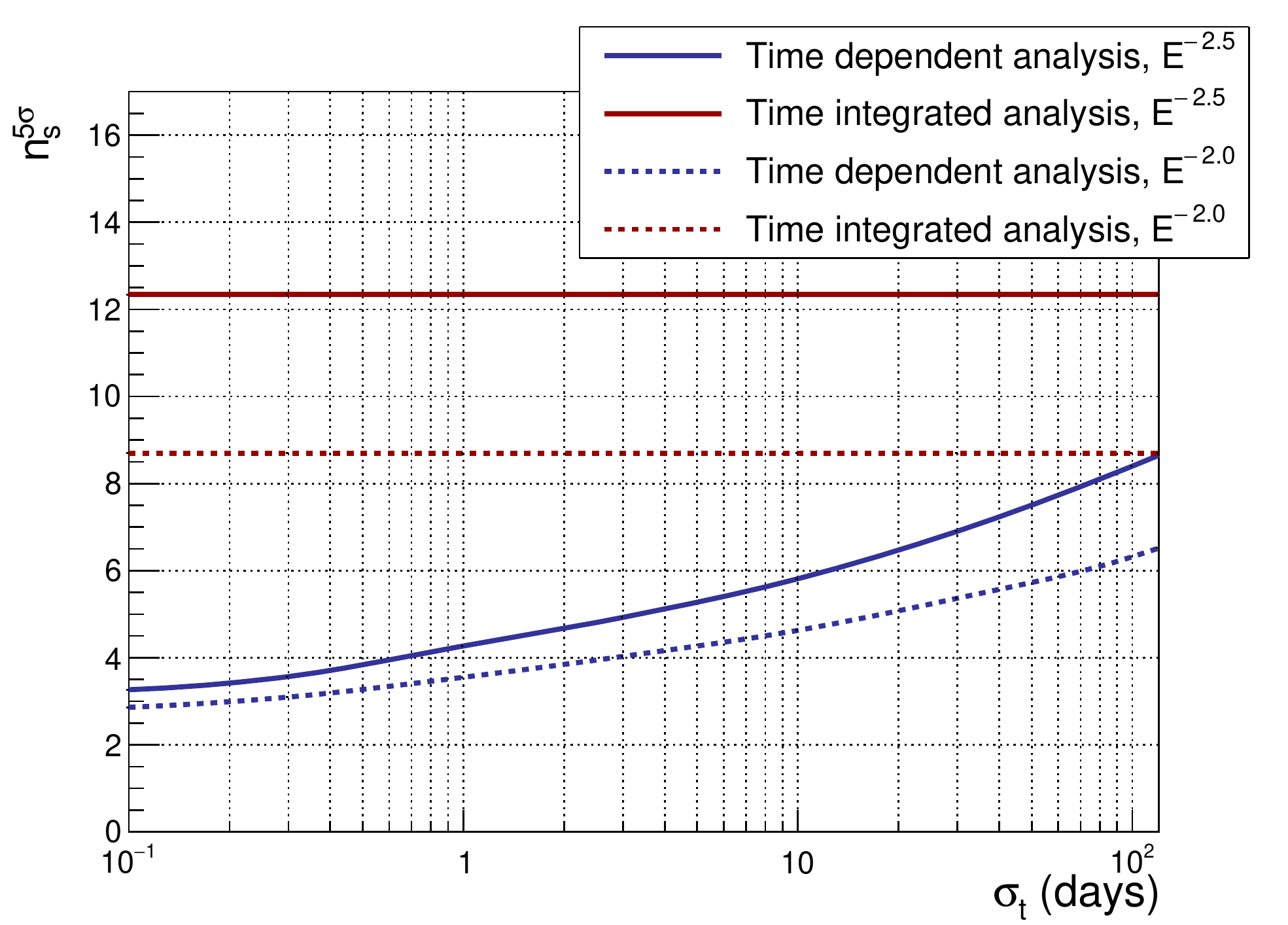}
    \caption{Mean number of signal events needed for a 5$\sigma$ discovery in 50\% of PEs for the ID 3 event of the IceCube HESE sample as a function of the flare duration $\sigma_t$. The result is shown for two assumptions of the energy spectrum: $E^{-2.5}$ (solid blue) and $E^{-2.0}$ (dotted blue). For comparison, the discovery potential of the time integrated analysis is also reported (red lines).}
    \label{fig:Discovery}
\end{figure}

In order to estimate the potential of the search, the mean number of signal events needed for a $5\sigma$ discovery is calculated for different durations of the simulated flare. As an example, Figure \ref{fig:Discovery} shows the number of signal events needed for a $5 \sigma$ significance with a 50\% detection power at the location of the IceCube event HESE ID 3, for a $E^{-\gamma}$ neutrino spectrum, with $\gamma$ equal to 2.0 or 2.5.

In the case of signal emission lasting a few hours, the number of events needed for a $5\sigma$ discovery is reduced by a factor $\sim 3$ (depending on the assumed source spectrum) with respect to a time integrated analysis. Similar levels of improvement in the discovery potential are expected for all the IceCube candidates.

\section{Results}\label{sec:RESULTS}

\begin{figure}[h]
    \centering
    \includegraphics[width=1.0\linewidth]{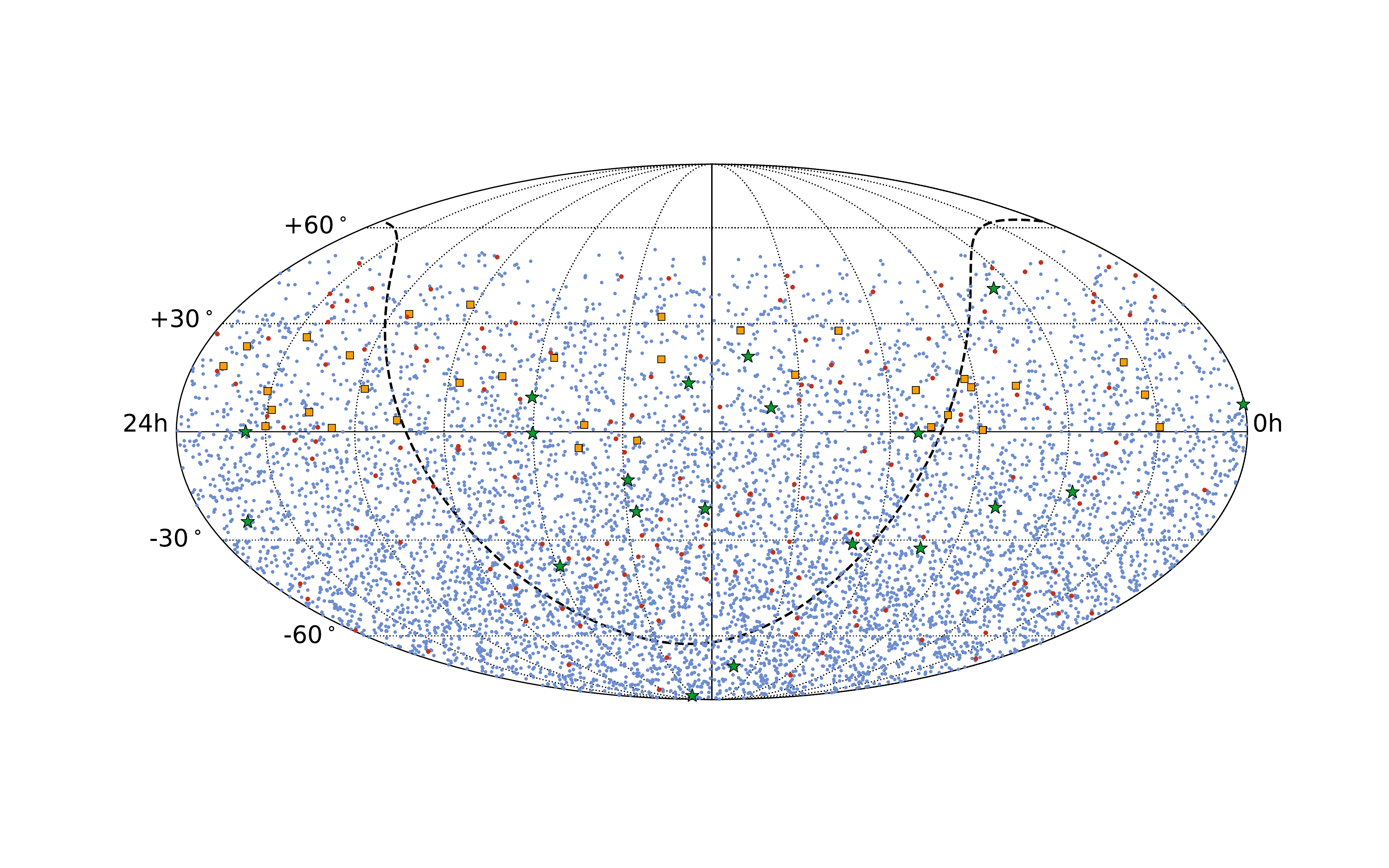}
    \caption{Sky map in equatorial coordinates of the 6894 track-like (blue circles) and the 160 shower-like (magenta circles) ANTARES events passing the selection cuts. Green stars and yellow squares show the location of the 20 and 34 neutrino candidates from the HESE and Muon IceCube samples, respectively. The black dashed line indicates the Galactic equator.}
    \label{fig:SkyMap}
\end{figure}

No significant excess over the expected background is observed for any of the assumed source locations when applying the described search method. The positions of the ANTARES tracks and showers together with the directions of the 54 IceCube candidates are shown in Figure \ref{fig:SkyMap}.

The most significant cluster, defined as the cluster with the lowest pre-trial p-value, is found at the location of the IceCube track with ID 15 from the Muon sample, with a number of fitted signal events $\hat{\mu}_\mathrm{sig} = 1.6$, a best-fit flare duration $\hat{\sigma}_t = 120$ days and a best-fit spectral index $\hat{\gamma} = 3.5$. 
The pre-trial p-value of the cluster is 3.7\%, corresponding to a significance of 2.1\,$\sigma$. The second and third most significant sources correspond to HESE ID 71 and Muon ID 26, with pre-trial p-values of 3.8\% and 4.6\%, respectively.
Since multiple candidates are analysed, trial factors must be taken into account. To do so, the distribution of the smallest p-values obtained performing the search on the same list of sources using PEs is computed. The observed pre-trial p-value is then compared to this distribution, providing a post-trial probability of 90\% for the most significant cluster. 
\newline In the absence of a significant excess, upper limits on the one-flavour neutrino fluence at 90\% C.L. are derived using the Neyman method \cite{neyman}. The fluence $\mathcal{F}$ is defined as the integral in time and energy of the neutrino energy flux $E \cdot \Phi_E$: 

\small
\begin{equation}
\label{eq:fluence}
\mathcal{F}  = \int_{t_{\rm min}}^{t_{\rm max}}\int_{E_{\rm min}}^{E_{\rm max}}E \cdot \Phi_E\ dE\ dt = \int_{t_{\rm min}}^{t_{\rm max}}\int_{E_{\rm min}}^{E_{\rm max}} E \cdot \Phi_0 \cdot S(E)\ dE\ dt = \Delta T \cdot \Phi_0 \cdot \int_{E_{\rm min}}^{E_{\rm max}} E \cdot S(E)\ dE  .
\end{equation}
\normalsize

In this equation, $\Phi_E$ is the neutrino differential flux given by the flux normalization $\Phi_0$ multiplied by the dimensionless neutrino spectrum $S(E) = (E/$GeV$)^{-\gamma}$. The integral in time extends over the duration of the flare $\Delta T$. In the integral~(\ref{eq:fluence}), the best-fit flare duration $\hat{\sigma_t}$ is assumed as $\Delta T$. The parameters $E_{\rm min}$ and $E_{\rm max}$ represent the boundaries of the declination-dependent energy range containing 90\% of the expected signal events.

\begin{figure*}[h] 
 	     \includegraphics[width=0.49\linewidth]{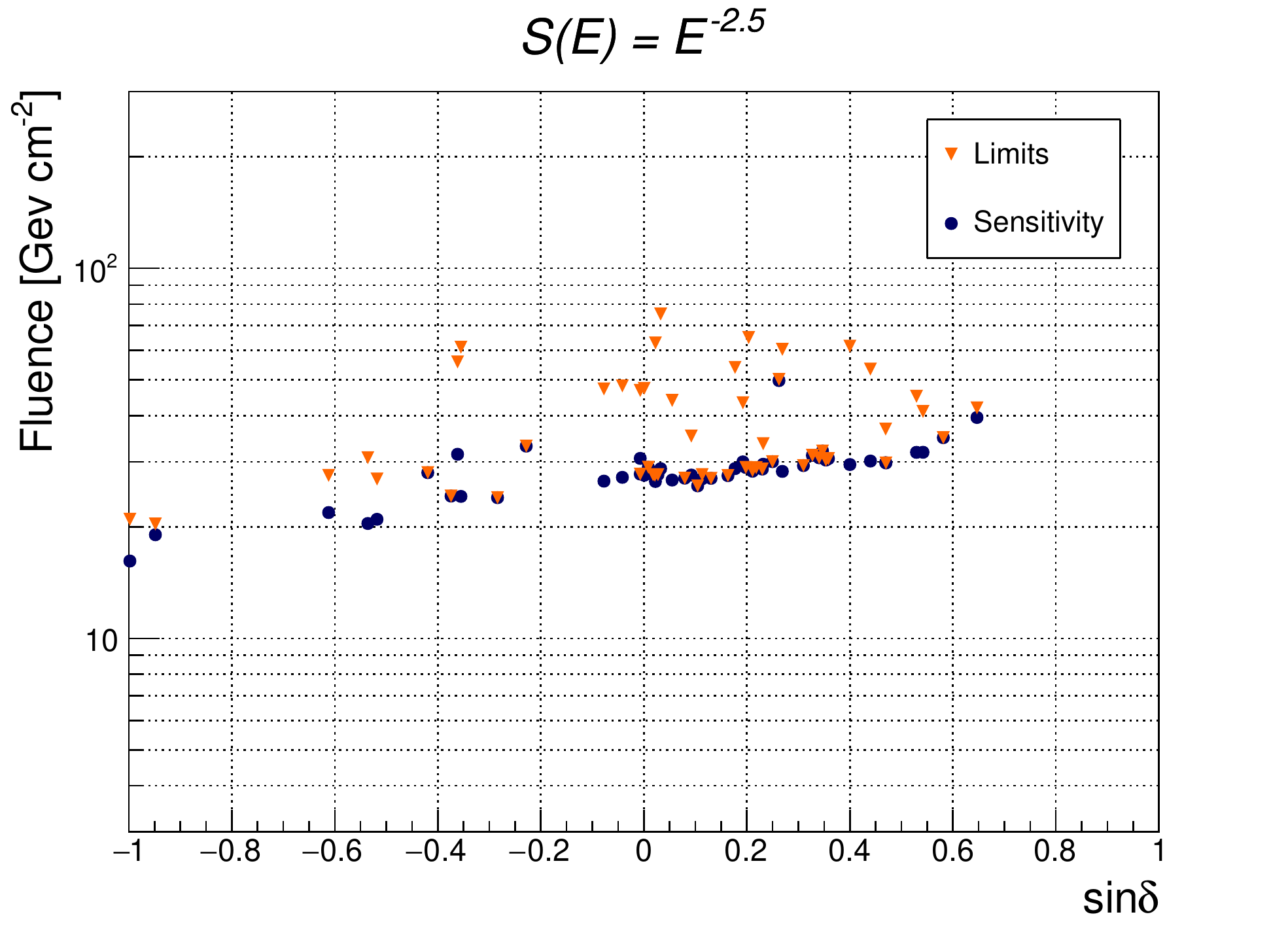}
       \includegraphics[width=0.49\linewidth]{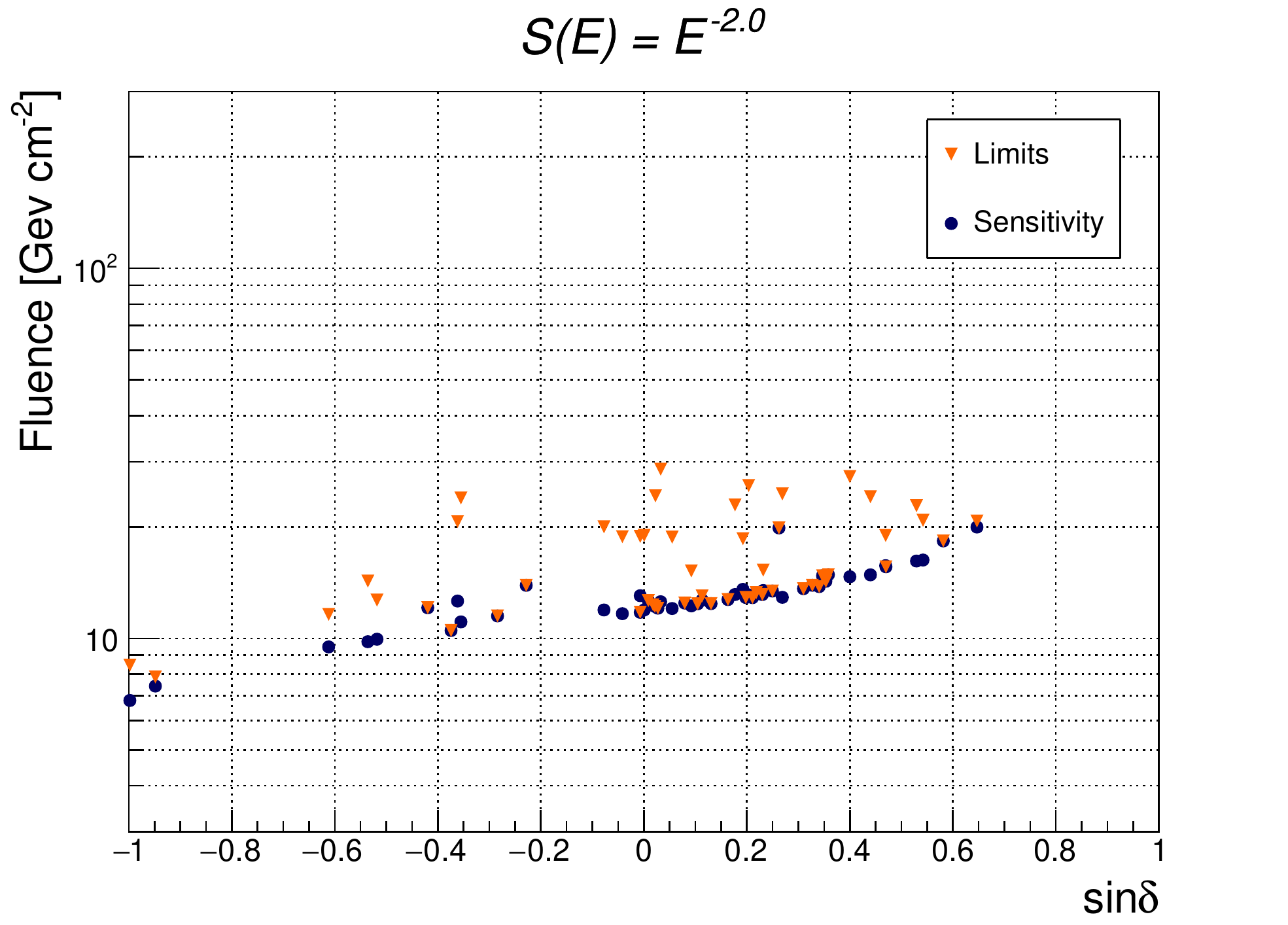}      
    \caption{Upper limits at $90\,\%$ C.L. on the one-flavour neutrino fluence (orange triangles) and sensitivities (blue dots) as a function of the investigated candidate declination for two assumptions of the signal energy spectrum: $S(E) = E^{-2.5}$ (left plot) and $S(E) = E^{-2.0}$ (right plot). Upper limits and sensitivities are calculated for the time windows reported in Tables~\ref{tab:candidatesHESE} and~\ref{tab:candidatesMuon}. A time window of $120\,\rm{days}$ is used for those sources with a null number of fitted signal events.}
    \label{fig:Limits}
\end{figure*}

A summary of the results, in terms of best-fit number of signal events $\hat{\mu}_\mathrm{sig}$, spectral index $\hat{\gamma}$, flare duration $\hat{\sigma}_t$ and upper limits on the fluence, is reported in Tables~\ref{tab:candidatesHESE} and~\ref{tab:candidatesMuon}. 
For those sources for which a null number of signal events is fitted,
limits are calculated assuming $\Delta T = 120\,\rm{days}$, chosen arbitrarily, as the value of the fitted flare duration is meaningless for clusters fully compatible with being background-like. In Figure~\ref{fig:Limits} the one-flavour neutrino fluence upper limits and sensitivities calculated for the same flares are shown as a function of the source declination for the two spectral assumptions. 

\setlength{\tabcolsep}{.3em}
\begin{table*}[p]
    \caption{\footnotesize List of analysed IceCube neutrino events from the HESE sample \cite{IC3years, IC4yproc, IC6yproc}. For each candidate, the equatorial coordinates -- declination ($\delta$) and right-ascension ($\alpha$) -- , date of observation, and angular error estimate $\beta_\mathrm{IC}$ are reported. The following four columns show the result of the search in terms of best-fit values for the likelihood function parameters (number of signal events $\hat{\mu}_\mathrm{sig}$, spectral index $\hat{\gamma}$, flare duration $\hat{\sigma}_t$) and $90\,\%$ C.L. upper limits on the one-flavour neutrino fluence for the two assumed energy spectral indices. Dashes (-) in the fitted likelihood parameters indicate sources with a null number of fitted signal events. The values of $E_{\rm min}$ and $E_{\rm max}$ used to calculate the fluence upper limits are listed in the last column.\medskip}
    \label{tab:candidatesHESE}
    \resizebox{\linewidth}{!}{
        \begin{minipage}{1.05\textwidth}
        \centering
 {\def\arraystretch{0.54}
    \footnotesize
    \begin{tabular}{c|c|c|c|c|c|c|c|c|c}
       \hline
      HESE ID  &$\delta [\si{\degree}]$     & $\alpha [\si{\degree}]$ & \makecell{observation time \\ $[$MJD$]$} & $\beta_\mathrm{IC} [\si{\degree}]$  & $\hat{\mu}_\mathrm{sig}$ & $\hat{\gamma}$ & $\hat{\sigma}_t [days]$ & \makecell{fluence limit [GeV cm$^{-2}$] \\ $\gamma = -2.5/-2.0$} & \makecell{log$(\frac{E_{\rm min}}{\textrm{GeV}})$ - log$(\frac{E_{\rm max}}{\textrm{GeV}})$ \\ $\gamma = -2.5/-2.0$}\\
\midrule   
  
3 & -31.2 & 127.9 & 55451.1 & 1.4 & 1.0 & 2.7 & 2.9 & 26.94 / 12.69 & 2.5 - 5.3 / 3.4 - 6.5 \\ 
5 & -0.4 & 110.6 & 55512.6 & 1.2 & 1.0 & 2.5 &  120  & 46.75 / 18.86 & 2.6 - 5.5 / 3.5 - 6.5 \\ 
8 & -21.2 & 182.4 & 55608.8 & 1.3 & 1.3 & 2.4 &  120  & 55.84 / 20.68 & 2.5 - 5.3 / 3.5 - 6.5 \\ 
13 & 40.3 & 67.9 & 55756.1 & 1.2 & 0.9 & 2.9 &  120  & 41.94 / 20.75 & 3.1 - 5.8 / 3.9 - 7.0 \\ 
18 & -24.8 & 345.6 & 55923.5 & 1.3 & - &  -  &  -  & 28.04 / 12.10 & 2.5 - 5.3 / 3.4 - 6.5 \\ 
23 & -13.2 & 208.7 & 55949.6 & 1.9 & 0.8 & 2.2 &  120  & 33.07 / 13.91 & 2.6 - 5.3 / 3.5 - 6.5 \\ 
28 & -71.5 & 164.8 & 56048.6 & 1.3 & 2.3 & 3.4 &  120  & 20.37 / 7.87 & 2.5 - 5.2 / 3.4 - 6.0 \\ 
37 & 20.7 & 167.3 & 56390.2 & 1.2 & - & - & - & 30.33 / 14.27 & 2.9 - 5.7 / 3.6 - 6.7 \\ 
43 & -22.0 & 206.6 & 56628.6 & 1.3 & 0.8 & 2.4 & 26.0 & 24.24 / 10.50 & 2.5 - 5.3 / 3.5 - 6.5 \\ 
44 & 0.0 & 336.7 & 56671.9 & 1.2 & 0.9 & 1.9 &  120  & 47.36 / 18.99 & 2.6 - 5.5 / 3.5 - 6.5 \\ 
45 & -86.2 & 219.0 & 56679.2 & 1.2 & 1.4 & 3.3 & 64.3 & 20.98 / 8.46 & 2.5 - 5.2 / 3.4 - 5.8 \\ 
53 & -37.7 & 239.0 & 56767.1 & 1.2 & 1.3 & 2.5 &  120  & 27.56 / 11.61 & 2.5 - 5.3 / 3.5 - 6.5 \\ 
58 & -32.4 & 102.1 & 56859.8 & 1.3 & 1.0 & 3.1 & 18.4 & 30.78 / 14.29 & 2.5 - 5.3 / 3.4 - 6.5 \\ 
61 & -16.5 & 55.6 & 56970.2 & 1.2 & - & - & - & 24.00 / 11.50 & 2.6 - 5.3 / 3.5 - 6.5 \\ 
62 & 13.3 & 187.9 & 56987.8 & 1.3 & - &  -  &  -  & 28.67 / 13.14 & 2.8 - 5.5 / 3.6 - 6.5 \\ 
63 & 6.5 & 160.0 & 57000.1 & 1.2 & 0.8 & 3.4 &  120  & 27.69 / 13.02 & 2.8 - 5.5 / 3.5 - 6.5 \\ 
71 & -20.8 & 80.7 & 57140.5 & 1.2 & 0.9 & 1.8 &  120  & 61.21 / 23.95 & 2.5 - 5.3 / 3.5 - 6.5 \\ 
76 & -0.4 & 240.2 & 57276.6 & 1.2 & - &  -  &  -  & 27.80 / 11.76 & 2.6 - 5.5 / 3.5 - 6.5 \\ 
78 & 7.5 & 0.4 & 57363.4 & 1.2 & - & - & - & 27.07 / 12.42 & 2.8 - 5.5 / 3.5 - 6.5 \\ 
82 & 9.4 & 240.9 & 57505.2 & 1.2 & - &  -  &  -  & 27.52 / 12.73 & 2.8 - 5.5 / 3.5 - 6.5 \\     

        \end{tabular}
      }
      \end{minipage}
      
    }
  
\end{table*}

\setlength{\tabcolsep}{.3em}
\begin{table*}[p]
    \caption{\footnotesize List of analysed IceCube neutrino events from the Muon sample \cite{IC-VHE-29, IC-VHE-36}. The same quantities as in Table~\ref{tab:candidatesHESE} are reported.\medskip}
    \label{tab:candidatesMuon}
    \resizebox{\linewidth}{!}{
        \begin{minipage}{1.05\textwidth}
        \centering
 {\def\arraystretch{0.54}
    \footnotesize
    \begin{tabular}{c|c|c|c|c|c|c|c|c|c}
       \hline
      Muon ID  &$\delta [\si{\degree}]$     & $\alpha [\si{\degree}]$ & \makecell{observation time \\ $[$MJD$]$} & $\beta_\mathrm{IC} [\si{\degree}]$  & $\hat{\mu}_\mathrm{sig}$ & $\hat{\gamma}$ & $\hat{\sigma}_t [days]$ & \makecell{fluence limit [GeV cm$^{-2}$] \\ $\gamma = -2.5/-2.0$} & \makecell{log$(\frac{E_{\rm min}}{\textrm{GeV}})$ - log$(\frac{E_{\rm max}}{\textrm{GeV}})$ \\ $\gamma = -2.5/-2.0$}\\
\midrule

1 & 1.2 & 29.5 & 55056.7 & 1.0 & - & - & - & 27.57 / 12.20 & 2.6 - 5.5 / 3.5 - 6.5 \\ 
2 & 11.7 & 298.2 & 55141.1 & 1.0 & 1.0 & 2.2 &  120  & 64.99 / 25.88 & 2.8 - 5.5 / 3.6 - 6.5 \\ 
3 & 23.6 & 344.9 & 55355.5 & 1.1 & 2.2 & 3.0 &  120  & 61.56 / 27.35 & 3.0 - 5.7 / 3.8 - 6.8 \\ 
5 & 21.0 & 307.0 & 55387.5 & 1.0 & - &  -  &  -  & 30.60 / 14.86 & 2.9 - 5.7 / 3.6 - 6.7 \\ 
6 & 15.2 & 252.0 & 55421.5 & 4.4 & 1.1 & 1.9 &  120  & 50.08 / 19.88 & 2.9 - 5.5 / 3.6 - 6.5 \\ 
7 & 13.4 & 266.3 & 55464.9 & 1.0 & 0.9 & 2.9 &  120  & 33.56 / 15.29 & 2.8 - 5.5 / 3.6 - 6.5 \\ 
8 & 11.1 & 331.1 & 55478.4 & 1.0 & 1.0 & 2.2 &  120  & 43.31 / 18.59 & 2.8 - 5.5 / 3.5 - 6.5 \\ 
9 & 0.5 & 89.0 & 55497.3 & 1.0 & 0.2 & 3.4 &  120  & 29.00 / 12.65 & 2.6 - 5.5 / 3.5 - 6.5 \\ 
10 & 3.1 & 285.9 & 55513.6 & 1.0 & 1.2 &  3.5  & 26.0 & 43.95 / 18.77 & 2.8 - 5.5 / 3.5 - 6.5 \\ 
11 & 1.0 & 307.7 & 55589.6 & 1.0 & - &  -  &  -  & 27.68 / 12.32 & 2.6 - 5.5 / 3.5 - 6.5 \\ 
12 & 20.3 & 235.1 & 55702.8 & 1.0 & - &  -  &  -  & 32.08 / 14.76 & 2.9 - 5.5 / 3.6 - 6.7 \\ 
13 & 35.5 & 272.2 & 55722.4 & 1.0 & - & - & - & 34.86 / 18.34 & 3.0 - 5.8 / 3.9 - 6.8 \\ 
14 & 5.3 & 315.7 & 55764.2 & 2.1 & 1.0 & 2.5 & 6.8 & 35.20 / 15.21 & 2.8 - 5.5 / 3.5 - 6.5 \\ 
15 & 1.9 & 222.9 & 55896.9 & 1.0 & 1.6 &  3.5  &  120  & 75.21 / 28.65 & 2.6 - 5.5 / 3.5 - 6.5 \\ 
16 & 19.1 & 36.6 & 55911.3 & 1.0 & - &  -  &  -  & 31.22 / 13.91 & 2.9 - 5.5 / 3.6 - 6.7 \\ 
17 & 32.0 & 198.7 & 56063.0 & 1.0 & 1.1 &  3.5  &  120  & 45.07 / 22.82 & 3.0 - 5.8 / 3.8 - 6.8 \\ 
18 & 1.6 & 330.1 & 56146.2 & 1.0 & 0.3 & 1.6 &  120  & 27.70 / 12.07 & 2.6 - 5.5 / 3.5 - 6.5 \\ 
19 & -2.4 & 205.1 & 56211.8 & 1.0 & 1.2 & 3.4 & 98.8 & 48.10 / 18.82 & 2.6 - 5.5 / 3.5 - 6.5 \\ 
20 & 28.0 & 169.6 & 56226.6 & 1.0 & - & - & - & 29.79 / 15.61 & 3.0 - 5.7 / 3.8 - 6.8 \\ 
21 & 14.5 & 93.4 & 56470.1 & 1.0 & - & - & - & 30.02 / 13.42 & 2.8 - 5.5 / 3.6 - 6.5 \\ 
22 & -4.4 & 224.9 & 56521.8 & 1.0 & 1.1 &  3.5  &  120  & 47.21 / 20.02 & 2.6 - 5.5 / 3.5 - 6.5 \\ 
23 & 10.2 & 32.9 & 56579.9 & 1.0 & 1.3 & 3.4 &  120  & 53.95 / 22.93 & 2.8 - 5.5 / 3.5 - 6.5 \\ 
24 & 32.8 & 293.3 & 56666.5 & 1.0 & 1.8 & 3.3 & 19.6 & 41.02 / 20.85 & 3.0 - 5.8 / 3.8 - 6.8 \\ 
25 & 18.1 & 349.4 & 56800.0 & 1.1 & - & - & - & 29.30 / 13.62 & 2.9 - 5.5 / 3.6 - 6.7 \\ 
26 & 1.3 & 106.3 & 56817.6 & 1.0 & 1.0 & 1.6 &  120  & 62.82 / 24.26 & 2.6 - 5.5 / 3.5 - 6.5 \\ 
27 & 11.4 & 110.6 & 56819.2 & 1.0 & - &  -  &  -  & 28.96 / 12.90 & 2.8 - 5.5 / 3.5 - 6.5 \\ 
28 & 4.6 & 100.5 & 57049.5 & 1.0 & - & - & - & 27.09 / 12.46 & 2.8 - 5.5 / 3.5 - 6.5 \\ 
29 & 12.2 & 91.6 & 57157.9 & 1.0 & - & - & - & 28.39 / 12.89 & 2.8 - 5.5 / 3.6 - 6.5 \\ 
30 & 26.1 & 325.5 & 57217.9 & 1.0 & 1.3 & 3.2 & 114.2 & 53.40 / 24.14 & 3.0 - 5.7 / 3.8 - 6.8 \\ 
31 & 6.0 & 328.4 & 57246.8 & 1.0 & - &  -  &  -  & 25.83 / 12.40 & 2.8 - 5.5 / 3.5 - 6.5 \\ 
32 & 28.0 & 134.0 & 57269.8 & 1.0 & 0.6 & 3.4 & 118.9 & 36.77 / 18.97 & 3.0 - 5.7 / 3.8 - 6.8 \\ 
33 & 19.9 & 197.6 & 57312.7 & 1.5 & - & - & - & 30.75 / 13.80 & 2.9 - 5.5 / 3.6 - 6.7 \\ 
34 & 12.6 & 76.3 & 57340.9 & 1.0 & - & - & - & 28.93 / 13.29 & 2.8 - 5.5 / 3.6 - 6.5 \\ 
35 & 15.6 & 151.3 & 57478.6 & 1.0 & 2.5 &  3.5  &  120  & 60.38 / 24.58 & 2.9 - 5.5 / 3.6 - 6.5 \\ 

        \end{tabular}
      }
      \end{minipage}
      
    }
  
\end{table*}

A discussion on the implications of the null observation in a time window of 0.1 days follows. The study does not reveal any ANTARES track-like (shower-like) event in correlation with any IceCube candidate within a time window of 0.1 days and a maximal angular distance of $\unit{10}{\degree}$ ($\unit{30}{\degree}$). Under the hypothesis that each IceCube candidate is produced by a different point-like transient source with a flare duration $\le$ 0.1 days, this non-detection is used to derive a constraint on the spectral index of such a source, as done in a previous ANTARES work \cite{ANTARESblazar}. Using a counting method and assuming Poisson statistics, the $90\%$ C.L. upper limit on the number of ANTARES events in time correlation with an IceCube HESE/Muon candidate is $n_s^{90} = 2.3$. The corresponding upper limit on the neutrino fluence normalisation for different spectral indices $\gamma$ is calculated as 

\begin{equation}
 \mathcal{F}_{\gamma}^{90} = \frac{n_s^{90}}{\int A^{\rm ANT}_{\rm eff}(E) \cdot E^{-\gamma}dE} ,
    \label{eq:limitfluence}
\end{equation}
where $A^{\rm ANT}_{\rm eff}$ is the ANTARES effective area.
The 90\% C.L. upper limit on the number of signal events expected to be observed by IceCube from a neutrino fluence $\mathcal{F}_{\gamma}^{90} E^{-\gamma}$ is then calculated as

\begin{equation}
N^{90}_{\nu , \rm IC} = \int \mathcal{F}_{\gamma}^{90} \cdot A_{\rm eff}^{\rm IC}(E) \cdot E^{-\gamma} dE ,
    \label{eq:limitIC}
\end{equation}
with $A_{\rm eff}^{\rm IC}$ being either the HESE or Muon IceCube effective area \cite{IceEff}.

\begin{figure}[h]
    \centering
    \includegraphics[width=0.65\linewidth]{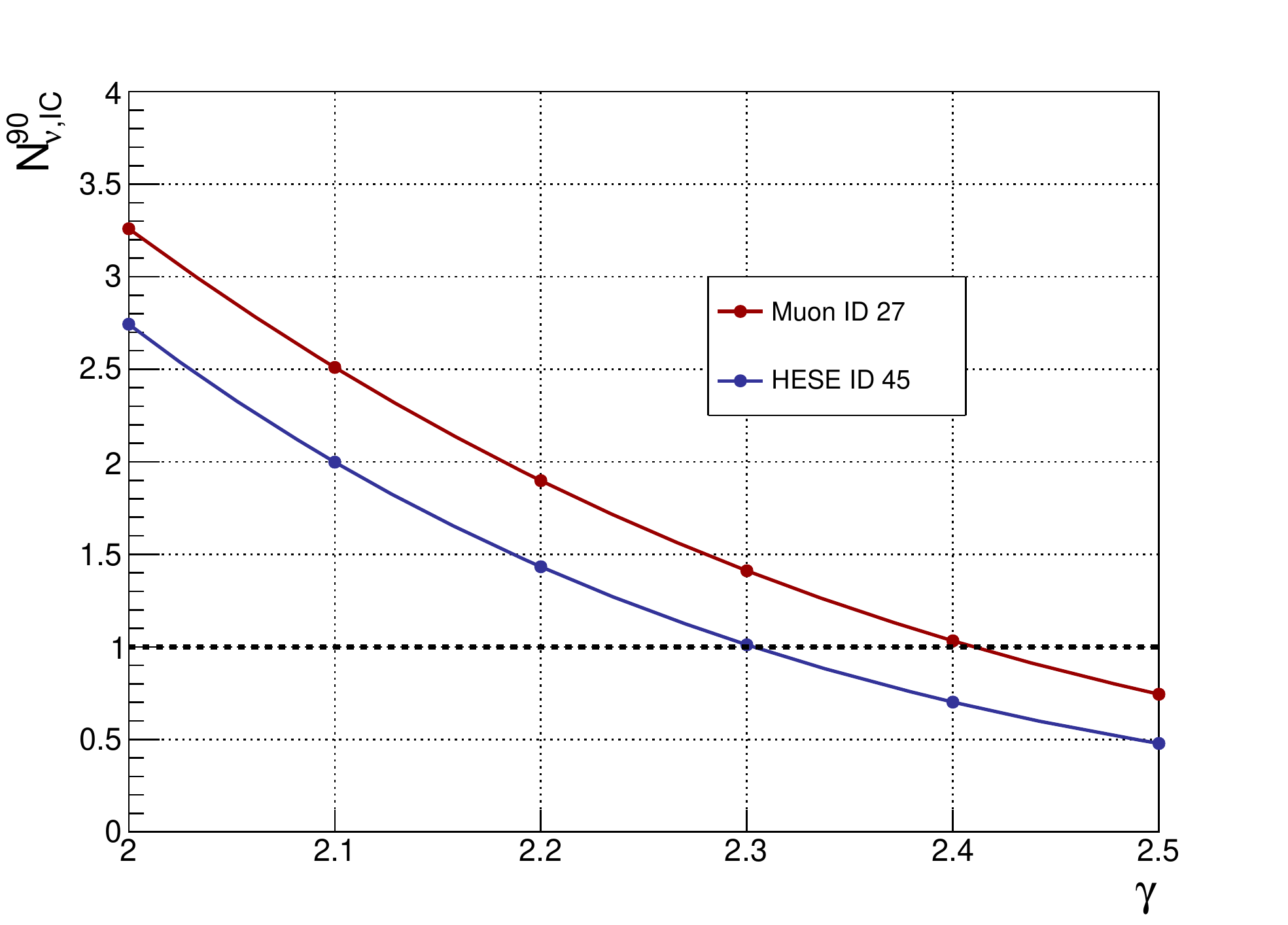}
    \caption{90\% C.L. upper limits on the expected number of IceCube events originated from a transient $E^{-\gamma}$ point-like source emitting in a time window $\le 0.1$ days as a function of the spectral index $\gamma$ for the most energetic IceCube event of the Muon sample, Muon ID 27 \cite{IC-VHE-29}, and of the HESE sample, HESE ID 45 \cite{IC4yproc}. The dotted line indicates the number of events detected by IceCube.}
    \label{fig:nevents}
\end{figure}

In Figure \ref{fig:nevents} the $90\%$ C.L. upper limits, $N^{90}_{\nu , \rm IC}$, as a function of the spectral index $\gamma$ are shown for the most energetic IceCube event of each sample, Muon ID 27 and HESE ID 45. If $N^{90}_{\nu , \rm IC}$ is smaller than 1 (number of events detected by IceCube), a transient origin with flare duration $\le 0.1$ days can be excluded at 90\% C.L.. Each IceCube event is therefore only consistent with the mentioned transient origin for neutrino spectra harder than $E^{-2.4}$ for the event Muon ID 27, $E^{-2.3}$ for the event HESE ID 45. These limits are compatible with the IceCube best-fitting spectral indices $2.2 \pm 0.2$ and $2.1 \pm 0.2$ for the neutrino flare from the direction of TXS 0506+056 \cite{ICTXS}.

\section{Conclusions}\label{sec:CONCL}

A search for time and space correlation between the selected ANTARES events and 54 IceCube high-energy track-like events has been presented. As no significant evidence of correlation between the ANTARES and the IceCube events has been observed, no further evidence is found to attribute the origin of the HESE and Muon neutrinos to a transient point-like source with flare duration between 0.1 and 120 days. Upper limits on the one-flavour neutrino fluence have been derived. The non-detection of any ANTARES event within 0.1 days from the IceCube neutrinos observation times has been used to constrain the spectral index of a possible flaring source responsible for the most energetic IceCube event of each sample.

\section*{Acknowledgements}
The authors acknowledge the financial support of the funding agencies:
Centre National de la Recherche Scientifique (CNRS), Commissariat \`a
l'\'ener\-gie atomique et aux \'energies alternatives (CEA),
Commission Europ\'eenne (FEDER fund and Marie Curie Program),
Institut Universitaire de France (IUF), IdEx program and UnivEarthS
Labex program at Sorbonne Paris Cit\'e (ANR-10-LABX-0023 and
ANR-11-IDEX-0005-02), Labex OCEVU (ANR-11-LABX-0060) and the
A*MIDEX project (ANR-11-IDEX-0001-02),
R\'egion \^Ile-de-France (DIM-ACAV), R\'egion
Alsace (contrat CPER), R\'egion Provence-Alpes-C\^ote d'Azur,
D\'e\-par\-tement du Var and Ville de La
Seyne-sur-Mer, France;
Bundesministerium f\"ur Bildung und Forschung
(BMBF), Germany; 
Istituto Nazionale di Fisica Nucleare (INFN), Italy;
Nederlandse organisatie voor Wetenschappelijk Onderzoek (NWO), the Netherlands;
Council of the President of the Russian Federation for young
scientists and leading scientific schools supporting grants, Russia;
Executive Unit for Financing Higher Education, Research, Development and Innovation (UEFISCDI), Romania;
Mi\-nis\-te\-rio de Econom\'{\i}a y Competitividad (MINECO):
Plan Estatal de Investigaci\'{o}n (refs. FPA2015-65150-C3-1-P, -2-P and -3-P, (MINECO/FEDER)), Severo Ochoa Centre of Excellence and Red Consolider MultiDark (MINECO), and Prometeo and Grisol\'{i}a programs (Generalitat
Valenciana), Spain; 
Ministry of Higher Education, Scientific Research and Professional Training, Morocco.
We also acknowledge the technical support of Ifremer, AIM and Foselev Marine
for the sea operation and the CC-IN2P3 for the computing facilities.


\providecommand{\href}[2]{#2}\begingroup\raggedright\endgroup

\end{document}